\documentstyle[12pt]{article}
\begin{document}
\def\be{\begin{equation}}
\def\ee{\end{equation}}
\def\Journal#1#2#3#4{{#1} {\bf #2}, #3 (#4)}
\def\NCA{\em Nuovo Cimento}
\def\NIM{\em Nucl. Instrum. Methods}
\def\NIMA{{\em Nucl. Instrum. Methods} A}
\def\NPB{{\em Nucl. Phys.} B}
\def\PLB{{\em Phys. Lett.}  B}
\def\PRL{\em Phys. Rev. Lett.}
\def\PRD{{\em Phys. Rev.} D}
\def\ZPC{{\em Z. Phys.} C}
\rightline{TECHNION-PH-99-17}
\vskip 3cm
\centerline{PENGUINS IN CP VIOLATING $B$ DECAYS
\footnote{Invited talk presented at
the 17th International Workshop on Weak Interactions and Neutrinos, Cape Town,
South Africa, January 24$-$30 1999}}
\bigskip\bigskip\bigskip
\centerline{Michael Gronau}
\bigskip
\centerline{\it Department of Physics}
\centerline{\it Technion -- Israel Institute of Technology, Haifa 32000,
Israel}

\vskip 4cm
\centerline{\bf ABSTRACT}
\medskip
\begin{quote}
The role of penguin amplitudes in CP violating $B$ decays
is reviewed, emphasizing recent progress in the analysis of electroweak penguin
contributions. It is shown how these terms are included in a model-independent
manner when
measuring the weak phase $\alpha$ in $B\to\pi\pi$ using isospin symmetry,
and when determining the phase $\gamma$ from $B\to K\pi$ applying flavor SU(3).
Uncertainties due to rescattering effects in $B\to K\pi$ are discussed.
\end{quote}
\newpage

\section{Introduction}
The long awaited recent report \cite{KTev} on a clear observation of direct CP
violation in $K\to\pi\pi$ decays, ${\rm Re}(\epsilon'/\epsilon) = (28.0 \pm
3.0 \pm 2.6 \pm 1.0)\times 10^{-4}$, is the first evidence for the important
role played by penguin amplitudes in the phenomena of CP violation
\cite{Paschos}. $B$ decays are expected to provide a variety of CP asymmetry
measurements, as well as measurerments of certain combinations of rates, some
of which carry the promise of determining the angles of the unitarity triangle
\cite{review}, $\alpha, \beta$ and $\gamma$. This can test the commonly
accepted hypothesis that CP violation arises solely from phases in the
Cabibbo-Kobayashi-Maskawa matrix \cite{KM}. Let us review \cite{talk} a few of
the ideas involved in this study, paying particular attention to the
role of penguin amplitudes.

\begin{itemize}
\item {\bf$\beta$}:
In the experimentally feasible \cite{CDF} and theoretically pure example of
$B^0(t)\to J/\psi K_S$ the decay amplitude is real to a very high
precision. Theoretically \cite{SanBi}, the time-dependent mixing-induced CP
asymmetry measures the phase $\beta\equiv -{\rm Arg}V_{td}$ controlling
$B^0$-$\bar B^0$ mixing to an accuracy of 1$\%$ \cite{pen}.
\item {\bf$\alpha$}:~
 $B^0(t)\to \pi^+\pi^-$ involves direct CP violation from the interference
between a dominant current-current amplitude carrying a weak phase $\gamma$
and a smaller penguin contribution, which ``pollutes" the measured $\sin\Delta
mt$ term in the time-dependent asymmetry \cite{pen}. A ratio of penguin to tree
amplitudes $|P/T|=0.3\pm 0.1$ in $B^0\to \pi^+\pi^-$ is inferred \cite{DGR}
from the measured rates \cite{CLEO} of $B\to K\pi$ dominated by a penguin
amplitude. Such a penguin contribution introduces a sizable uncertainty
\cite{MG} in the determination of $\alpha=\pi-\beta-\gamma$ in
$B^0\to\pi^+\pi^-$. Isospin
symmetry may be used \cite{GL} to remove this unknown correction to $\alpha$ by
measuring also the time-integrated rates of $B^{\pm}\to\pi^{\pm}\pi^0$ and
$B^0(\bar B^0)\to \pi^0\pi^0$. In the likely case that the decay rate into
$\pi^0\pi^0$ cannot be measured with sufficient precision, one can at least
use this measurement to set upper limits on the error in $\alpha$ \cite{GQ}.
Further out in the future, one may combine the time-dependence of $B^0(t)\to
\pi^+\pi^-$ with the U-spin related $B_s(t)\to K^+K^-$
to determine separately $\beta$ and $\gamma$ \cite{Duni}. This involves
uncertaities due to SU(3) breaking.
\item {\bf$\gamma$}:
The angle $\gamma$ is apparently the most difficult to measure.
It was suggested some time ago \cite{GLR} to obtain information about this
angle from charged $B$ decays to $K\pi$ final states by measuring the
relative phase between a dominant real penguin amplitude and a smaller
current-current amplitude carrying the phase $\gamma$. This is achieved by
relating the latter amplitude through flavor SU(3) \cite{GHLR} to the amplitude
of $B^+\to\pi^+\pi^0$, introducing SU(3) breaking in terms of $f_K/f_{\pi}$.
\end{itemize}

In the above two examples of determining $\alpha$ and $\gamma$, QCD penguin
amplitudes were taken into account in terms of their very general properties,
whereas electroweak penguin (EWP)
contributions were first neglected and later on analyzed in a model-dependent
manner \cite{ewp}. Such an approach relies on factorization and on
form factor assumptions \cite{model}, and involves theoretical uncertainties in
hadronic matrix elements similar to those plaguing
$\epsilon'/\epsilon$ \cite{Paschos}.

In the present report we will focus on recent developments in the
study of EWP contributions, which partially avoid these
uncertainties, thereby improving the potential accuracy of measuring
$\alpha$ and $\gamma$.

\section{Model-independent treatment of electroweak penguins}

The weak Hamiltonian governing $B$ decays is given by \cite{Buras}

\be\label{H}
{\cal H} = \frac{G_F}{\sqrt2}
\sum_{q=d,s}\left(\sum_{q'=u,c} \lambda_{q'}^{(q)}
[c_1 Q_1 + c_2 Q_2]
-  \lambda_t^{(q)}\sum_{i=3}^{10}c_i Q^{(q)}_i\right)~,
\ee
where $Q_1=(\bar bq')_{V-A}(\bar q'q)_{V-A}, Q_2=(\bar bq)_{V-A}
(\bar q'q')_{V-A},~\lambda_{q'}^{(q)}=V_{q'b}^*V_{q'q},~q=d,s,~q'=u,c,t,
~\lambda_u^{(q)}+\lambda_c^{(q)}+\lambda_t^{(q)}=0$.
The dominant EWP operators $Q_9,~Q_{10}$ ($|c_{7,8}|\ll |c_{9,10}|$)
have a (V-A)(V-A) chiral structure, similar to the current-current operators
$Q_1, Q_2$. Thus, isospin alone
relates the matrix elements of these operators in $B^+\to \pi^+\pi^0$
\cite{GPY}
\be\label{EWpi}
\sqrt2 P^{EW}(B^+\to \pi^+\pi^0)=\frac32 \kappa(T+C)~,~~~~~\kappa=\frac{c_9 +
c_{10}}{c_1 + c_2} = -0.0088~,
\ee
where $T+C$ represents graphically \cite{GHLR} the current-current amplitudes
dominating $B^+\to \pi^+\pi^0$. Similarly, flavor
SU(3) implies \cite{GPY}
\be\label{EW1}
P^{EW}(B^+\to K^0\pi^+) + \sqrt2 P^{EW}(B^+\to K^+\pi^0) =
\frac32\kappa (T+C)~,
\ee
\be\label{EW2}
P^{EW}(B^0\to K^+\pi^-) + P^{EW}(B^+\to K^0\pi^+) =
\frac32 \kappa (C-E)~.
\ee
In the next three sections we describe briefly applications of these three
relations to the determination of
$\alpha$ and $\gamma$ from $B\to \pi\pi$ and $B\to K\pi$, respectively.

\section{Controlling EWP contributions in $B\to\pi\pi$}

The time-dependent rate of $B^0\to \pi^+\pi^-$ includes a term
$\sim\sin(2\alpha+
\theta)\sin(\Delta mt)$, where the correction $\theta$ is due to penguin
amplitudes \cite{GL}. Using isospin (\ref{EWpi}), the EWP
contribution to $\theta$, denoted by $\xi$, is found to be very small
\cite{GPY,BF}
\be
\tan\xi = \frac{x\sin\alpha}{1+x\cos\alpha},~~
x\equiv \frac32\kappa
|\frac{\lambda_t^{(d)}}{\lambda_u^{(d)}}| = -0.013
|\frac{\lambda_t^{(d)}}{\lambda_u^{(d)}}|~,
\ee
and is nicely incorporated into the analysis of Ref.~12 which determines
$\alpha$.

\section{$\gamma$ from $B^+\to K\pi$}

Using (\ref{EW1}), EWP terms are included in the triangle construction of
Ref.~15  \cite{NR2}
\be
\sqrt2 A(B^+\to K^+\pi^0) + A(B^+\to K^0\pi^+) =
\tilde r_u A(B^+\to\pi^+\pi^0) \left(1 - \delta_{EW} e^{-i\gamma}\right)~,
\ee
where $\tilde r_u = (f_K/f_{\pi})\tan\theta_c\simeq 0.28,~
\delta_{EW}=-(3/2)|\lambda^{(s)}_t/\lambda^{(s)}_u|\kappa \simeq 0.66\pm 0.15$.
This relation and its charge-conjugate permit a determination of $\gamma$
\cite{GLR,NR2}
under the {\it assumption} that a rescattering amplitude with phase
$\gamma$ can
be neglected in $B^+\to K^0\pi^+$. This amplitude is bounded
by the U-spin related rate of $B^{\pm}\to K^{\pm}\bar K^0$ \cite{Falk,GR,FL}.
Present limits are at the level of $20-30\%$ of the dominant penguin amplitude
\cite{GPY,Neubert}, and are expected to be improved to the level of 10$\%$.
In this case the rescattering effect, which depends strongly on
the final state phase difference $\phi$ between $I=3/2$ current-current and
penguin amplitudes,
introduces an uncertainty at a level of $15^{\circ}$ in the determination of
$\gamma$ if $\phi$ is near $90^{\circ}$ \cite{GP}.  A considerably smaller
theoretical error \cite{Neubert} would be implied if this measurable phase is
found to be far
from $90^{\circ}$.

Other sources of errors in $\gamma$, such as SU(3) breaking, are
discussed elsewhere at this meeting \cite{Neubert,Flei}.
We note that in this determination of $\gamma$ SU(3) breaking does not occur
in the leading penguin amplitudes as it does in some other methods \cite{Duni}.

The phase $\gamma$ can also be constrained by measuring only charge-averaged
$B^{\pm}\to
K\pi$ rates. Defining
\be\label{R*def}
R^{-1}_*=\frac{2[B(B^+\to K^+\pi^0) + B( B^-\to K^-\pi^0)]}
{B(B^+\to K^0\pi^+) + B(B^-\to \bar K^0\pi^-)}~,
\ee
one finds using (\ref{EW1}) \cite{GPY,NR1}
\be
R^{-1}_* = 1 - 2\epsilon \cos\phi (\cos\gamma - \delta_{EW}) +
{\cal O}(\epsilon^2, \epsilon^2_A, \epsilon\epsilon_A)~,
\ee
where \cite{GLR,NR1} $\epsilon = \tilde r_u \sqrt 2
|A(B^{\pm}\to\pi^{\pm}\pi^0)/A(B^{\pm}\to
K^0\pi^{\pm})|\sim 0.24$, while $\epsilon_A$ is the suitably normalized
rescattering
amplitude. The resulting bound
\be\label{const}
|\cos\gamma - \delta_{EW}| \ge \frac{|1-R^{-1}_*|}{2\epsilon}~,
\ee
which neglects {\it second order} corrections, can be used to exclude an
interesting region around $\cos\gamma = \delta_{EW}$ if $R^{-1}_*\ne 1$ is
measured. Again, this would be very difficult if $\phi\simeq 90^{\circ}$. The
present value of the ratio of  rates is \cite{CLEO} $R^{-1}_*=2.1\pm 1.1$.

\section{$\gamma$ from the ratio of $B^0\to K^{\pm}\pi^{\mp}$ to $B^{\pm}\to
K^0\pi^{\pm}$ rates}

Denoting this ratio of charged-averaged rates by $R$ ~\cite{FM},
one finds using
(\ref{EW2}) a constraint very similar to (\ref{const}) \cite{GPY,BF,FL}
\be\label{const'}
|\cos\gamma - \delta'_{EW}| \ge \frac{|1-R|}{2\epsilon'}~
\ee
where $\delta'_{EW}\sim 0.2\delta_{EW}\sim 0.13$ represents color-suppressed
EWP contributions, and \cite{GR}~ $\epsilon'\sim 0.2$ is the ratio of tree to
penguin amplitudes in $B^0\to K^+\pi^-$. In contrast to (\ref{const}), this
bound neglects {\it first order} rescattering effects, and the values of
$\delta'_{EW}$ and $\epsilon'$ are less solid than those of $\delta_{EW}$ and
$\epsilon$ in (\ref{const}). Eq.~(\ref{const'})  can exclude a region around
$\gamma=90^{\circ}$ if $R\ne 1$ is found. Presently \cite{CLEO} $R=1.07\pm
0.45$.

\section{Conclusion}
\begin{itemize}
\item In $B\to\pi\pi$ strong and electroweak penguins are controlled by isospin.
\item In $B\to K\pi$ strong penguins dominate and EWP are controlled by SU(3).
\item Interesting bounds on $\gamma$, in one case susceptible to rescattering
effects, are implied if the $B\to K\pi$ charge-averaged ratios of rates differ
from 1.
\item A precise determination of $\gamma$ from $B\to K\pi$ is challenging and
requires a combined effort involving further theoretical and experimental
studies.
\end{itemize}

\noindent
{\bf Acknowledgment}:
This work is supported by the United States $-$
Israel Binational Science Foundation under Research Grant Agreement 94-00253/3.



\begin{thebibliography}{99}
\bibitem{talk} This is a very short summary of a 45 minute review talk
presented
at this conference. Relevant references are included for further details.
\end{thebibliography}
\end{document}